\documentclass[twocolumn]{aastex631f}

\shorttitle{Fibrils}
\shortauthors{Bate et al.}
\graphicspath{{./}{}}

\usepackage{gensymb}
\usepackage{amsmath}
\usepackage{subfigure}
\DeclareMathOperator{\sech}{sech}

\begin{document}

\title{Unveiling the True Nature of Plasma Dynamics from the Reference Frame of a Super-penumbral Fibril}

\correspondingauthor{W.~Bate}
\email{wbate02@qub.ac.uk}

\author[0000-0001-9629-5250]{W.~Bate}
\affiliation{Astrophysics Research Centre, School of Mathematics and Physics, Queen's University Belfast, Belfast, BT7 1NN, UK}

\author[0000-0002-9155-8039]{D.~B.~Jess}
\affiliation{Astrophysics Research Centre, School of Mathematics and Physics, Queen's University Belfast, Belfast, BT7 1NN, UK}
\affiliation{Department of Physics and Astronomy, California State University Northridge, 18111 Nordhoff Street, Northridge, CA 91330, USA}

\author[0000-0001-5170-9747]{S.~D.~T.~Grant}
\affiliation{Astrophysics Research Centre, School of Mathematics and Physics, Queen's University Belfast, Belfast, BT7 1NN, UK}

\author[0000-0002-0851-5362]{A.~Hillier}
\affiliation{Department of Mathematics and Statistics, University of Exeter, Exeter, EX4 4QF, UK}

\author[0000-0002-3814-4232]{S.~J.~Skirvin}
\affiliation{ Centre for mathematical Plasma Astrophysics, Mathematics Department, KU Leuven, Celestijnenlaan 200B bus 2400, B-3001 Leuven, Belgium}

\author[0000-0001-9628-4113]{T.~van~Doorsselaere}
\affiliation{ Centre for mathematical Plasma Astrophysics, Mathematics Department, KU Leuven, Celestijnenlaan 200B bus 2400, B-3001 Leuven, Belgium}

\author[0000-0002-7711-5397]{S.~Jafarzadeh}
\affiliation{Max Planck Institute for Solar System Research, Justus-von-Liebig-Weg 3, 37077 G{\"o}ttingen, Germany}
\affiliation{Niels Bohr International Academy, Niels Bohr Institute, Blegdamsvej 17, DK-2100 Copenhagen, Denmark}

\author[0000-0001-6238-0721]{T.~Wiegelmann}
\affiliation{Max Planck Institute for Solar System Research, Justus-von-Liebig-Weg 3, 37077 G{\"o}ttingen, Germany}

\author[0000-0003-3306-4978]{T.~Duckenfield}
\affiliation{Astrophysics Research Centre, School of Mathematics and Physics, Queen's University Belfast, Belfast, BT7 1NN, UK}

\author[0000-0001-7706-4158]{C.~Beck}
\affiliation{National Solar Observatory, 3665 Discovery Drive, Boulder, CO 80303, USA}

\author[0000-0001-8385-3727]{T. Moore}
\affiliation{Astrophysics Research Centre, School of Mathematics and Physics, Queen's University Belfast, Belfast, BT7 1NN, UK}

\author[0000-0002-5365-7546]{M.~Stangalini}
\affiliation{ASI, Italian Space Agency, Via del Politecnico snc, 00133, Rome, Italy}

\author[0000-0001-8556-470X]{P.~H. Keys}
\affiliation{Astrophysics Research Centre, School of Mathematics and Physics, Queen's University Belfast, Belfast, BT7 1NN, UK}

\author[0000-0003-1746-3020]{D.~J. Christian}
\affiliation{Department of Physics and Astronomy, California State University Northridge, 18111 Nordhoff Street, Northridge, CA 91330, USA}




\begin{abstract}

The magnetic geometry of the solar atmosphere, combined with projection effects, makes it difficult to accurately map the propagation of ubiquitous waves in fibrillar structures. These waves are of interest due to their ability to carry energy into the chromosphere and deposit it through damping and dissipation mechanisms. To this end, the Interferometric Bidimensional Spectrometer (IBIS) at the Dunn Solar Telescope was employed to capture high resolution H$\alpha$ spectral scans of a sunspot, with the transverse oscillations of a prominent super-penumbral fibril examined in depth. The oscillations are re-projected from the helioprojective-cartesian frame to a new frame of reference oriented along the average fibril axis through non-linear force-free field extrapolations. The fibril was found to be carrying an elliptically polarised, propagating kink oscillation with a period of $430$~s and a phase velocity of $69\pm4$~km{\,}s$^{-1}$. The oscillation is damped as it propagates away from the sunspot with a damping length of approximately $9.2$~Mm, resulting in the energy flux decreasing at a rate on the order of $460$~W{\,}m$^{-2}$/Mm. The H$\alpha$ line width is examined and found to increase with distance from the sunspot; a potential sign of a temperature increase. Different linear and non-linear mechanisms are investigated for the damping of the wave energy flux, but a first-order approximation of their combined effects is insufficient to recreate the observed damping length by a factor of at least $3$. It is anticipated that the re-projection methodology demonstrated in this study will aid with future studies of transverse waves within fibrillar structures.

\end{abstract}



\section{Introduction} 
\label{sec:intro}

\begin{figure*}[t!]
\includegraphics[trim=30mm 0mm 5mm 0mm, clip, width=\textwidth, angle=0]{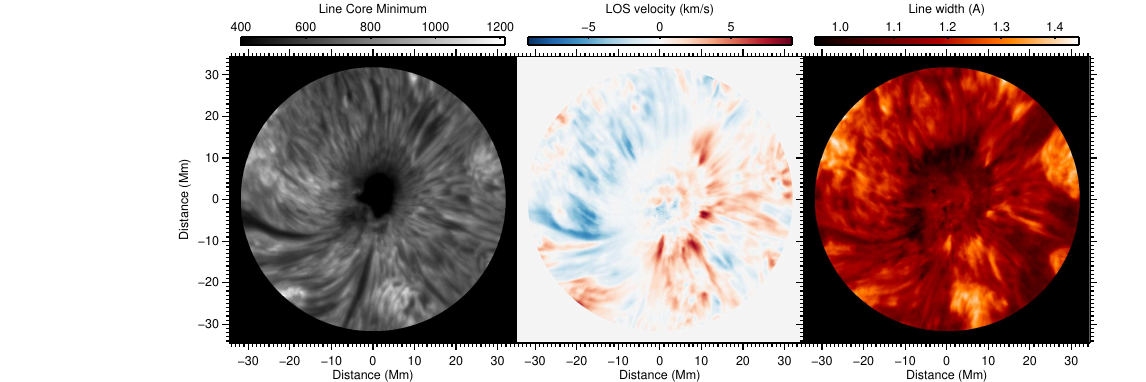}
 \caption{Data products derived from the IBIS field-of-view, acquired at $14$:$51$~UT on 2015 September 16. The left panel shows the H$\alpha$ line core minimum found using the fit described by Equation~\ref{eqn:Super-sech}. The middle panel displays the derived line of sight velocity from the parameter $A_{1}$ in Equation~\ref{eqn:Super-sech}. The right panel shows the associated H$\alpha$ line width in units of {\AA}ngstr\"{o}ms ({\AA}).
\label{fig:FOV}}
\end{figure*}

Solar fibrils are thin, elongated structures observed in the chromosphere of the Sun, appearing as dark, thread-like features that trace the magnetic field lines above the solar surface \citep{2017ApJS..229...11J}. Though prominent in regions of strong magnetic activity, such as sunspots or active regions, they are in-fact ubiquitous across the solar disk. They are believed to be the manifestation of magnetic field-aligned plasma density enhancements and are closely related to other solar features such as spicules. These similar features are often grouped under the umbrella term ``fibrillar structures'' \citep[see, e.g., the recent review by][]{2023LRSP...20....1J}.

The exact physical processes that give rise to solar fibrils are still a subject of ongoing investigation \citep{2012SSRv..169..181T}. It is thought that they form due to the interaction of the plasma with the magnetic field lines, trapping denser plasma into field-aligned structures, giving the appearance of elongated structures \citep{2015ApJ...802..136L}. What is clear is they are most abundant in the thin and diffuse chromosphere \citep{2012NatCo...3.1315M, 2014RAA....14..299L, Jess2015}. In this dynamic region, fibrils have the potential to facilitate the transfer of energy between the photosphere and corona \citep{2014ApJ...784...29M}. Thus, they are a vital component to studies of energy transfer and deposition in the solar atmosphere since they act as visual tracers for wave phenomena.

Transverse magnetohydrodynamic (MHD) kink waves have been the subject of study in small-scale structures within sunspots. \citet{2011ApJ...739...92P} detected kink waves in dynamic fibrils near a sunspot, with a period of $135$~s, using Ca~{\sc{ii}} $854.2$~nm observations from the Swedish Solar Telescope (SST). More recently, \citet{2021RSPTA.37900183M} employed high-resolution observations in the same spectral line from SST/CRISP to demonstrate the presence of transverse waves in sunspot super-penumbral fibrils, located in the solar chromosphere. They interpreted these oscillations as MHD kink modes, with average periods and propagation speeds of approximately $754$~s and $25$~km{\,}s$^{-1}$, respectively. The velocity amplitudes, with an average of $0.76\pm0.47$~km{\,}s$^{-1}$, were observed to increase by about $80\%$ with increasing distance from the centre of the sunspot. \citet{2021RSPTA.37900183M} speculated that this variation might be due to a decrease in density along the fibrils as the super-penumbral region extends to higher atmospheric heights while moving away from the umbra, until it reaches the highest point of the magnetic canopy and returns to the surface. Hence, considering the field geometry in the chromosphere becomes crucial when interpreting these observations, particularly in intensity images where projection effects are not readily apparent. \citet{2021RSPTA.37900183M} also discussed several potential excitation mechanisms for the transverse oscillations, such as convection-driven motions, magnetic reconnection, and mode conversion, ultimately finding the mode conversion mechanism to be the most convincing. It is worth noting that oscillations in these small-scale structures within sunspots may differ from those observed in other chromospheric features due to factors such as the exceptionally strong magnetic fields of a few kG present in sunspots \citep{2006SoPh..239...41L}.

\citet{Jafarzadeh2017c} examine transverse oscillations of bright Ca~{\sc{ii}}~H fibrils in the vicinity of a forming sunspot with a similar magnetic topology to that of a sunspot. Median periods and velocity amplitudes of $83\pm29$~s and $2.4\pm0.8$~km~s$^{-1}$ respectively, for oscillations within $134$ fibrils. Propagating waves were able to be characterised in $23$ fibrils, with median propagation speeds of $9\pm14$~km~s$^{-1}$. Propagation in a single direction within each fibril was the most common occurrence, although counter-propagating waves within a single fibril, as well as standing waves, were also observed.

Finding the energy flux of fibril oscillations is an important step in investigating their contribution to the heating of the chromosphere and corona. It is estimated that an energy flux of $10^{3}-10^{4}$~W{\,}m$^{-2}$ is required to heat the chromosphere \citep{Withbroe1977}, an order of magnitude more than coronal requirements. It should be emphasised that this energy must be transported into the chromosphere and also dissipated there, i.e., it is not sufficient for the energy to simply travel through. This suggests that accounting for chromospheric heating is a challenge of equal or greater magnitude than for coronal heating when investigating solar atmospheric heating mechanisms.

An important caveat when interpreting any transverse energy flux estimates is that they are only based on resolved oscillations. Waves with amplitudes too small to be spatially resolved or periods too short to be temporally resolved are not included in these estimations, leaving the possibility that a significant amount of wave energy may be unaccounted for \citep{2016GMS...216..431V}. Another aspect contributing to the underestimation of the total energy flux may be the presence of kink motions along the observer's line-of-sight, which will not manifest as visible transverse oscillations. Examples of this have been documented by \citet{2018ApJ...853...61S} and \citet{2021ApJ...921...30S}, who measured helical motions of spicules through Doppler measurements \citep[see also the modeling work by][]{2008ApJ...683L..91Z}. If these line-of-sight motions are not taken into account when calculating the energy flux, it may result in an underestimation of the true value \citep{2021ApJ...922...60S}. Further, superposition of multiple features along the line-of-sight can result in an underestimation of the wave energy calculated from Doppler velocity oscillations by $1-3$ orders of magnitude \citep{2012ApJ...746...31D,2019ApJ...881...95P}.

In previous work, \citet{2022ApJ...930..129B} examined transverse kink oscillations within off-limb H$\alpha$ spicules using imaging observations and found that the energy flux of the upwardly propagating transverse waves decreases with atmospheric height at a rate of $-13{\,}200\pm6{\,}500$~W{\,}m$^{-2}$/Mm. As a result, this decrease in energy flux as the waves propagate upwards may provide significant thermal input into the local plasma.

The aim of the current study is to expand on previous work on the energy flux of transverse waves within fibrillar structures. We aim to examine the properties of transverse oscillations of super-penumbral fibrils within three-dimensional space as they travel along the structures, allowing for the probing of their energy carrying capabilities and any damping that may occur. Further, we aim to take advantage of the properties of the H$\alpha$ line in order to probe the plasma temperature and search for signs of MHD wave energy dissipation. To achieve this objective, we utilize ground-based instrumentation with high spatial, spectral, and temporal resolutions, providing data products that are ideally suited for this study. 

\section{Observations} 
\label{sec:obs}

Observations were taken on 2015 September 16 with the Interferometric Bidimensional Spectrometer \citep[IBIS;][]{2006SoPh..236..415C} at the Dunn Solar Telescope (DST) of the decaying active region (AR) NOAA 12418. The data were utilised by \citet{Beck2020} and \citet{Prasad2022}, where they are discussed in detail, thus only a short description is presented here. The IBIS field of view was placed over the leading sunspot of the group, located at $[x,y]=[-520'',-340'']$ with respect to solar disk centre. The IBIS field of view consists of a circular aperture with a diameter of $95''$ and a spatial sampling of $0{\,}.{\!\!}''095$ per pixel. IBIS acquired data from $14$:$42-15$:$16$~UT, with H$\alpha$ spectral scans of $27$ non-equidistant wavelength points. The exposure time at each wavelength position was $40$~ms, with a full-scan cadence of $11.2$~s, resulting in $400$ scans being taken with the majority under good seeing. All IBIS scans were co-aligned by correlating subsequent images of the continuum intensity with each other. These co-alignment corrections were minimal in nature due to the use of advanced adaptive optics \citep[AO;][]{Rimmele2011} utilised at the DST. The left panel of Figure~\ref{fig:FOV} shows the H$\alpha$ line-core intensity captured at $14$:$51$~UT. 

A single vector magnetogram was acquired at $14$:$46$:$19$~UT using the Helioseismic and Magnetic Imager \citep[HMI;][]{2012SoPh..275..327S} on board the Solar Dynamics Observatory \citep[SDO;][]{2012SoPh..275....3P}. This magnetogram was obtained with a spatial scale of $0{\,}.{\!\!}''5$ per pixel. The full disk magnetogram was cropped to only include AR~NOAA $12418$ using the associated Space-weather HMI Active Region Patch \citep[SHARP;][]{2014SoPh..289.3549B}. This was used to create a magnetic field extrapolation, detailed in Section~\ref{sub:reproj}.

\section{Data Analysis} 
\label{sec:ana}
In the following sub-sections, we detail extensively the methodology employed to accurately fit the H$\alpha$ spectral line (Section~{\ref{sub:line}}), isolate the fibrillar feature of interest (Section~{\ref{sub:fibrilfit}}), and crucially re-project the fibril dynamics into a common reference frame based around its central axis (Section~{\ref{sub:reproj}}). These steps are fundamentally important to ensure robust consistency with future studies of fibrillar activity in mind.

\subsection{Line Fitting}
\label{sub:line}

\begin{figure}[t]
\centering
\includegraphics[trim=0mm 0mm 0mm 0mm, clip, width=\columnwidth, angle=0]{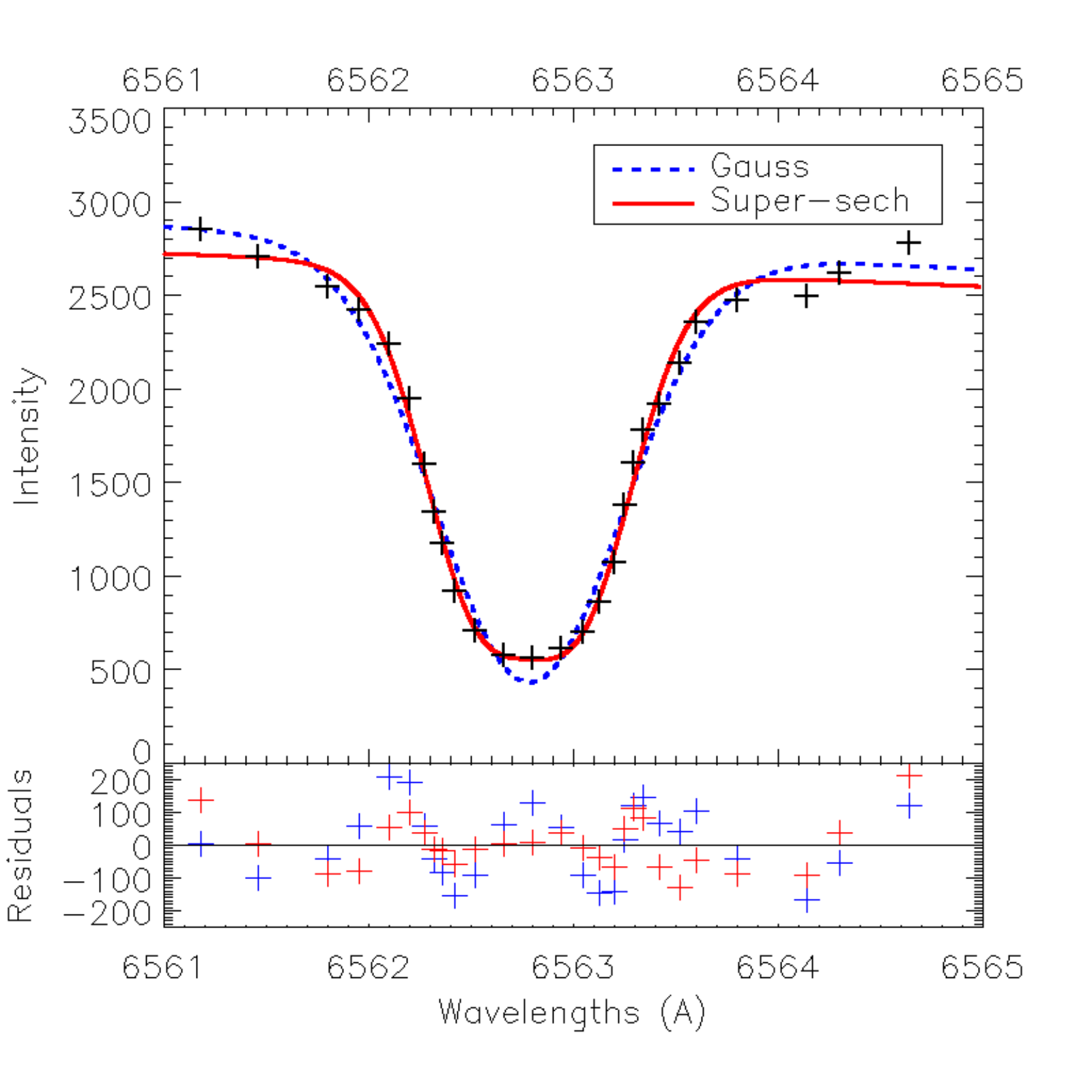}
 \caption{A demonstration of the line fitting routines used on the H$\alpha$ profile taken from a pixel within a fibril. In the upper panel, the pluses show the IBIS data; the dashed blue line shows a Gaussian profile fit described by equation~\ref{eqn:Gauss}; and the solid blue line shows a super-sech profile fit described by equation~\ref{eqn:Super-sech}. The lower panel shows the residuals of these two fits with the same colours. 
\label{fig:sech}}
\end{figure}

In a solar context, the H$\alpha$ absorption line is known to be broadened by a number of factors, including turbulence, non-thermal, and thermal effects. The line shape is also affected by the fine and hyperfine structure of the $3\rightarrow2$ electronic transition. In addition, the line is also sensitive to strong magnetic fields, with an effective Land\'{e} factor of $1.06$ \citep{doi:10.1080/14786441108564938, 1982SoPh...77..285L, 2018MNRAS.477.2796L}. This leads to the line shape not being fit well by typical line profiles such as the Gaussian, Lorentzian, or Voigt \citep{2012ApJ...749..136L}. To illustrate this, a simple Gaussian fit of an example H$\alpha$ profile is shown in blue in Figure~\ref{fig:sech}, described by
\begin{equation}
\label{eqn:Gauss}
    f_{G}(\lambda)=-A_{0}\exp\bigg(-\frac{(\lambda-A_{1})^{2}}{2A_{2}^{ 2}}\bigg) + Q(\lambda) \ ,
\end{equation} 
where $A_{0}$ denotes the line depth, $A_{1}$ corresponds to the wavelength shift of the line core position, $A_{2}$ is a parameter that is linearly related to the line-width, and $Q$ is a quadratic trend to aid with the fitting of the continuum.

\begin{figure}[t]
\centering
\includegraphics[trim=0mm 0mm 0mm 0mm, clip, width=\columnwidth, angle=0]{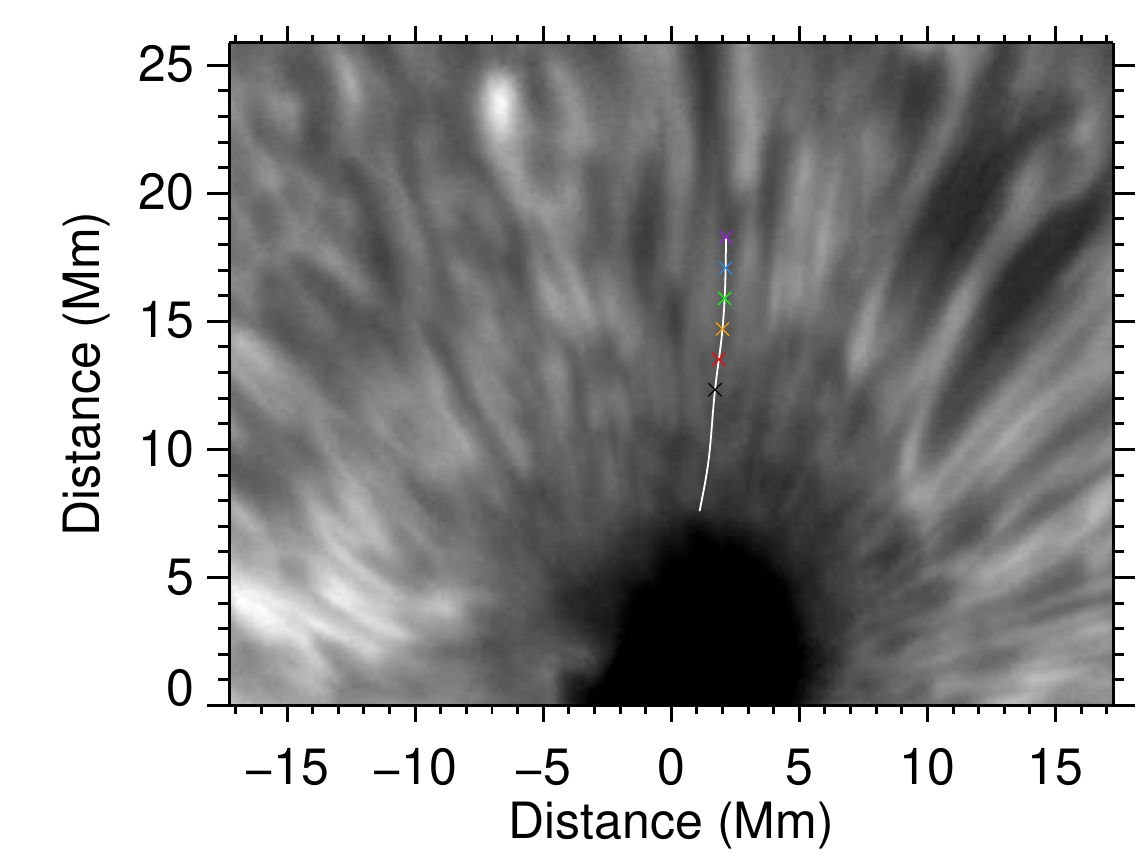}
 \caption{The example fibril analysed in Section~\ref{sec:res} is shown overplotted as a white line on a line core H$\alpha$ intensity image. The location of the cross cuts that were taken are marked by coloured crosses, with the colours corresponding to those in Figure~\ref{fig:temp}. 
\label{fig:fibril}}
\end{figure}

It can be seen that the simple Gaussian profile demonstrated in Figure~{\ref{fig:sech}} systematically underestimates the line-core intensity and does not well represent the line width, with a consistent underestimation towards the line centre and a consistent overestimation further out into the line wings. 

As semi-automated analysis of the line shape, and especially the line width, was desired, many other potential functions to describe the H$\alpha$ line profile were tested using a chi-squared goodness of fit test. The function that performed best is described as follows,

\begin{equation}
\label{eqn:Super-sech}
    f_{SS}(\lambda)=-A_{0}\sech^{2}\Bigg(\bigg(\frac{(\lambda-A_{1})^{2}}{2A_{2}^{ 2}}\bigg)^{A_{3}}\Bigg) + Q(\lambda)\ ,
\end{equation}
where $A_{3}$ is a parameter describing the steepness of the sides of the line profile between the core and continuum, and $A_{0,1,2}$ and $Q$ have similar meanings to those in Equation~\ref{eqn:Gauss}. The function described by equation~\ref{eqn:Super-sech} will henceforth be referred to as `super-sech' due to its similarity to a higher order Gaussian (also known as a super-Gaussian). The mean-squared deviation (MSD) averaged over all line profiles in the dataset was found to be $4{\,}170$ for the super-sech profile, compared to an MSD of $12{\,}990$ for a simple Gaussian and $5{\,}250$ for a higher order (or super-) Gaussian. This clearly demonstrates the suitability of Equation~\ref{eqn:Super-sech} over Equation~\ref{eqn:Gauss} for fitting the H$\alpha$ line profiles within this dataset.

The prior estimates for the super-sech parameters ($A_{0,1,2}$) are taken from an initial fit using a simple Gaussian profile to reduce computational time. This is possible as the estimation of the central wavelength shift ($A_{1}$) is consistent between the Gaussian and super-sech fits, hence a Gaussian fit remains valid for any studies investigating line of sight velocities. This super-sech fitting routine will be made publicly available with the release of WaLSAtools\footnote{WaLSAtools website: \href{https://walsa.team/codes}{https://WaLSA.team/codes}}, anticipated for later this year \citep{2023LRSP...20....1J}. Line of sight velocity ($v_{\text{LOS}}$) and line width ($W_{H\alpha}$) maps are shown in the central and right panels of Figure~\ref{fig:FOV}, respectively. Here, the line of sight velocity is calculated using the parameter $A_{1}$ in equation~\ref{eqn:Super-sech}.

The H$\alpha$ line width is known to correlate with the plasma temperature \citep[e.g.,][]{2012ApJ...749..136L,2019ApJ...881...99M,2023A&A...673A.137P}. Hence, the precision of the super-sech method of H$\alpha$ line fitting allows for temperature inference through the derived methods described by \citet{2009A&A...503..577C}. This method of line width calculation was chosen for ease of comparison to the work of \citet{2019ApJ...881...99M} and use of their derived relationship, 

\begin{equation}
\label{eqn:tempwidth}
    W_{H\alpha}=6.12\times 10^{-5} \times T + 0.533 \ ,
\end{equation}
where $W_{H\alpha}$ is the width of the H$\alpha$ line, and $T$ is the brightness temperature observed in Atacama Large Millimeter Array \citep[ALMA;][]{2009IEEEP..97.1463W} Band $3$. This ALMA brightness temperature represents the electron temperature at the formation height of the formation height of the $3$~mm radiation \citep{2015A&A...575A..15L, 2016SSRv..200....1W}.  Only wavelengths within $\pm1.0$~{\AA} of the nominal line core wavelength were used in the calculation of the line width as this mirrors the methodology used in \citet{2019ApJ...881...99M}. Care must be taken when using this relationship, however, as the exact slope can vary depending on the solar structure imaged \citep{2023FrASS...978405T}.

\subsection{Fibril Fitting}
\label{sub:fibrilfit}

Individual fibrils were identified by examination of H$\alpha$ line core intensity images averaged over the length of the dataset. This allows us to identify the fibrils that persist for a large proportion of the time which is a useful condition for seeing many oscillation cycles, and further allows for the identification of their average axis position. An example fibril was chosen to best demonstrate the techniques used in the analysis of fibril oscillations. The calculated average fibril axis for this object was fit using cubic splines. Cross-cuts were then taken at multiple points along the fibril, with each cut perpendicular to the fibril axis in a similar mannner to the methodology utilised by \citet{2021RSPTA.37900183M}. The example fibril, further discussed in Section~\ref{sec:res}, is shown in Figure~\ref{fig:fibril} with the locations of the cross cuts taken denoted by coloured crosses.

This has allowed for the production of time-distance (TD) diagrams of line core intensity ($I_{C})$, line of sight velocity ($v_{\text{LOS}}$), and line width ($W_{H\alpha}$) for each cut across the fibril, examples of which are shown in Figure~\ref{fig:td}, taken at the position of the green cross in Figure~\ref{fig:fibril}. The transverse deviations of the fibril from its average axis can be traced by fitting Gaussians to each time slice of the $I_{C}$ TD diagrams in a method similar to that of \citet{Morton2013} \& \citet{Weberg2018}. These transverse motions are overplotted in each panel of Figure~\ref{fig:td}. The $v_{\text{LOS}}$ and $W_{H\alpha}$ values can then be read using the calculated transverse positions as a probe.

\subsection{Re-projection to Fibril Axis Frame}
\label{sub:reproj}

Many previous works on the motion of fibrillar structures have utilised a Helioprojective-Cartesian (HPC) co-ordinate system \citep{2006A&A...449..791T}. This is due to the ease of use when considering observations taken from only a single point and can be easily derived from the detector (CCD) co-ordinates. This approach, however, can `flatten' two main sources of projection effects, losing information about the fibrillar structure's motion in the process. These main projection effects are from (1)~The angle between the solar normal and the observer's line-of-sight (the cosine of which is typically referred to as~$\mu$), and (2)~The angle between the axis of the fibrillar structure and the solar surface. In order to better discuss the three-dimensional motion of fibrils a new co-ordinate system defined in terms of the fibril axis has been utilised, henceforth referred to as the Fibril Axis Frame (FAF), and the process of its derivation and use is described below.

\begin{figure}[!t]
     \centering

         \includegraphics[trim=0mm 0mm 0mm 0mm, clip, width=\columnwidth, angle=0]{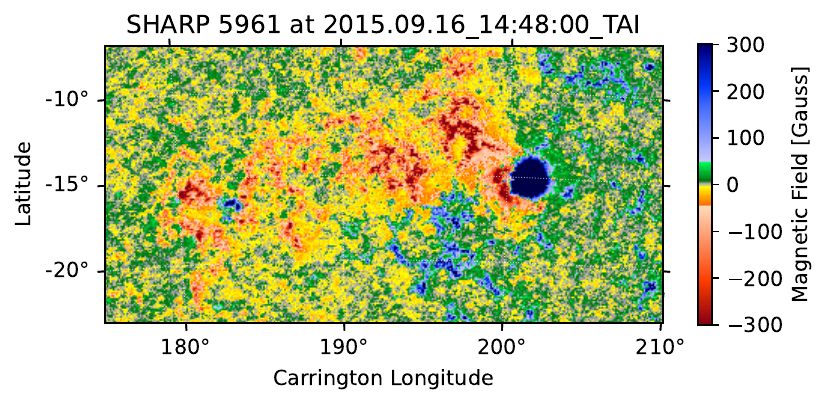}
     \hfill

         \includegraphics[trim=0mm 0mm 0mm 0mm, clip, width=\columnwidth, angle=0]{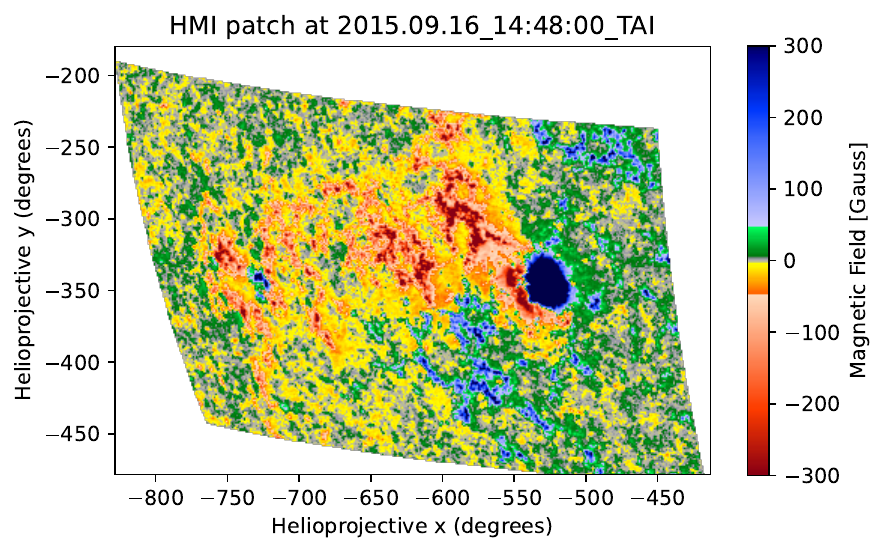}
     \hfill

        \caption{The top panel shows the HMI SHARP $5961$, used for the NLFFF extrapolation. This is shown with the associated CEA co-ordinates. The lower panel shows this same patch used as a mask on HMI data and the associated helioprojective co-ordinates. The magnetic field in the upper panel is calculated with respect to an observer directly above the sunspot centre, in the lower panel it is with respect to the line of sight of SDO. Both magnetic field values are cropped at $\pm300$~G for clarity. }
        \label{fig:reproj}
\end{figure}

A Non-Linear Force-Free Field \citep[NLFFF;][]{2005A&A...433..701W, 2008JGRA..113.3S02W} extrapolation was performed in order to understand the angle of the fibril axis with respect to both the solar surface and the observer's line of sight \citep[see][for a recent review of solar force-free fields]{2021LRSP...18....1W}. This is achievable as fibrils typically trace magnetic field lines \citep{2012ApJ...749..136L}, and the penumbral region of sunspots exist in a low plasma-$\beta$ state \citep{Gary2001, Grant2018}. This was achieved using data from the Helioseismic and Magnetic Imager \citep[HMI;][]{2012SoPh..275..207S} as well as the Space-weather HMI Active Region Patch \citep[SHARP;][]{2014SoPh..289.3549B} for SHARP $5961$, associated with AR NOAA $12418$ shown in the top panel of Figure~\ref{fig:reproj}. The NLFFF extrapolation was performed utlising the Lambert Cylindrical Equal-Area (CEA) projection. The original vector magnetogram has been preprocessed in order to make it consistent with a force-free field in the corona \citep[see][for details]{2006SoPh..233..215W}. The NLFFF extrapolations have been carried out with the help of an optimization code. The force-free optimization principle has been described in \citet{2000ApJ...540.1150W,2004SoPh..219...87W} and the NLFFF-code version specified for SDO/HMI (which we used here) in \citet{2012SoPh..281...37W}.

\begin{figure}[t]
\centering
\includegraphics[trim=5mm 60mm 100mm 10mm, clip, width=\columnwidth, angle=0]{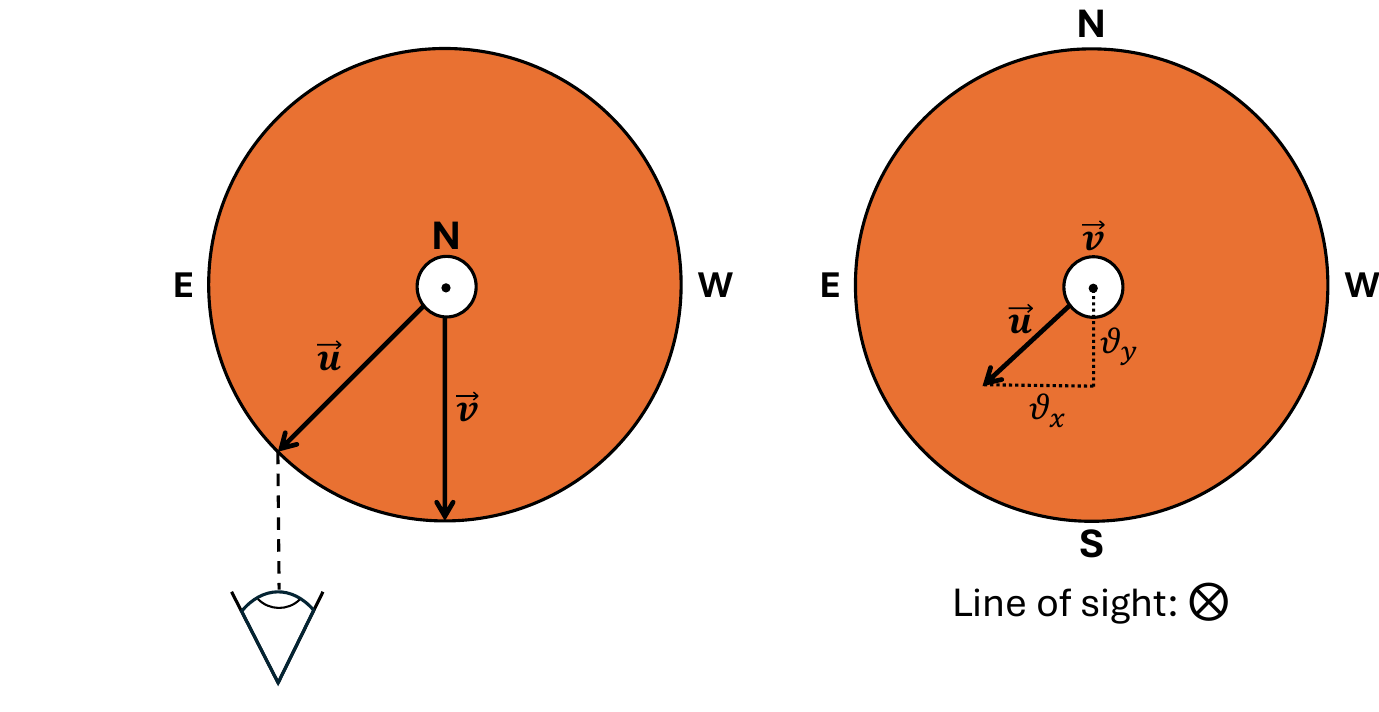}
 \caption{ A schematic diagram to indicate the directions of the vectors $\vec{\mathbf{u}}$ and $\vec{\mathbf{v}}$, described by Equations~\ref{eqn:u} and \ref{eqn:v} respectively. The left panel shows a top-down view of the Sun and the right panel shows the view from the direction of the Earth. Here $\bigodot$ represents a vector pointing out of the page, and $\bigotimes$ represents a vector pointing into the page.
\label{fig:uv_vectors}}
\end{figure}

The magnetic fields derived from the NLFFF model were re-projected into HPC co-ordinates in order to interpret the plane-of-sky and line-of sight oscillations. This is a two step process. First, the positions of each magnetic field vector were translated from CEA to HPC co-ordinates using SunPy's coordinates module \citep{sunpy_community2020}, following a similar approach to that of \citet{2013arXiv1309.2392S}. The resulting area covered in HPC co-ordinates is shown in the lower panel of Figure~\ref{fig:reproj}. Second, the magnetic field vectors were then transformed with the aid of a rotation matrix, $\mathcal{R}$. This was constructed for a transformation from the vector $\vec{\mathbf{u}}$, which points from the solar centre and through the centre of the feature being examined in HPC co-ordinates, to the vector $\vec{\mathbf{v}}$, which points from the solar centre to the solar surface along the observer's line of sight. Hence, $\vec{\mathbf{u}}$ and $\vec{\mathbf{v}}$ are given by,
\begin{equation}
\label{eqn:u}
    \vec{\mathbf{u}} = \begin{pmatrix}
    \vartheta_{x} & \vartheta_{y} & h
    \end{pmatrix}
    ^{T} \ ,
    \ \hat{\mathbf{u}} = \frac{\vec{\mathbf{u}}}{| {\vec{\mathbf{u}}} |} 
    \ ,
\end{equation}
\begin{equation}
\label{eqn:v}
    \vec{\mathbf{v}} = \begin{pmatrix}
    0 & 0 & \vartheta_{limb}
    \end{pmatrix}
    ^{T} \ ,
    \ \hat{\mathbf{v}} = \frac{\vec{\mathbf{v}}}{| {\vec{\mathbf{v}}} |} 
    \ ,
\end{equation}
where $\vartheta_{x}$ and $\vartheta_{y}$ are the $x$ and $y$ CD co-ordinates of the feature under investigation, $\vartheta_{limb}$ is the distance from solar centre to the limb in CD co-ordinates, and ${h=\sqrt{\vartheta_{limb}^{2} - \sqrt{\vartheta_{x}^{2} + \vartheta_{y}^{2}}}}$. The directions of $\vec{\mathbf{u}}$ and $\vec{\mathbf{v}}$ are shown in Figure~\ref{fig:uv_vectors}.

\begin{figure}[t]
\centering
\includegraphics[trim=50mm 45mm 40mm 0mm, clip, width=\columnwidth, angle=0]{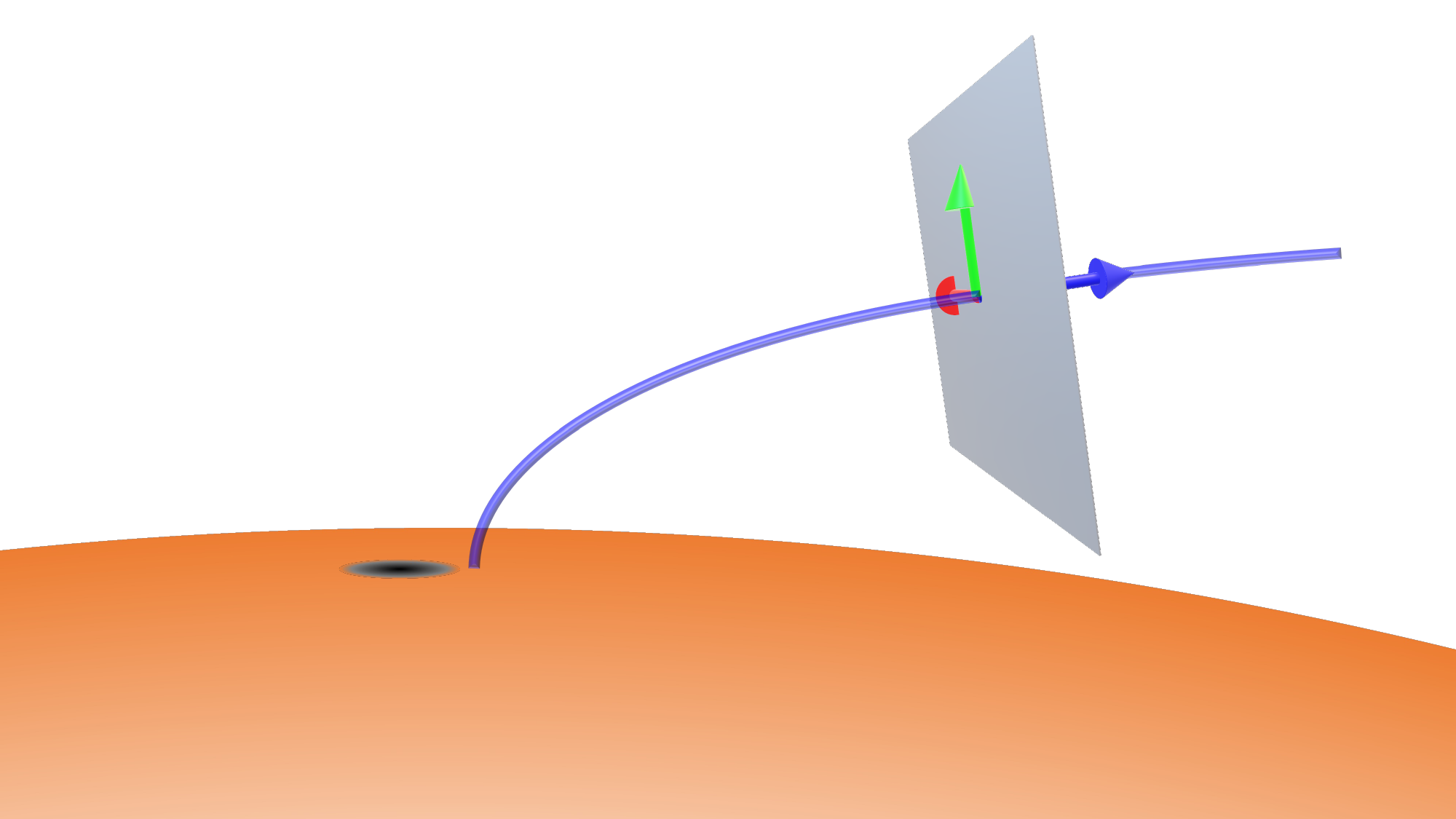}
 \caption{A cartoon of the Fibril Axis Frame (FAF). The fibril axis is marked by the blue curved tube. The FAF-$x$, $y$, and $z$ axes are shown by the red, green, and blue arrows respectively. Note that the $x$ axis is parallel to the solar surface, and the $z$ axis is co-directional with the fibril axis and becomes more positive further from the sunspot. The FAF-$x,y$ plane is shown by the grey square.  
\label{fig:fib_cart}}
\end{figure}

A mutually perpendicular vector, $\vec{\mathbf{s}}$, to both $\vec{\mathbf{u}}$ and $\vec{\mathbf{v}}$, was constructed as,
\begin{equation}
\label{eqn:s}
    \vec{\mathbf{s}} = \frac{\hat{\mathbf{u}} \times \hat{\mathbf{v}}} {|\hat{\mathbf{u}} \times \hat{\mathbf{v}}|} 
    \ ,
    \ \hat{\mathbf{s}} = \frac{\vec{\mathbf{s}}}{| {\vec{\mathbf{s}}} |} = 
    \begin{pmatrix}
    s_{0} & s_{1} & s_{2}
    \end{pmatrix}
    ^{T} 
    \ .
\end{equation}
It should be noted that due to the construction of $\vec{\mathbf{u}}$ and $\vec{\mathbf{v}}$, $s_{2}$ is always equal to $0$. This allows the construction of our rotation matrix,
\begin{equation}
\label{eqn:R}
   \mathcal{R} = \begin{pmatrix}
    \cos{\theta} + s_{0}^{2}(1-\cos{\theta}) & s_{0}s_{1}(1-\cos{\theta}) & s_{1}\sin{\theta} \\
    s_{1}s_{0}(1-\cos{\theta}) & \cos{\theta} + s_{1}^{2}(1-\cos{\theta}) & -s_{0}\sin{\theta} \\
    -s_{1}\sin{\theta} & s_{0}\sin{\theta} & \cos{\theta}
    \end{pmatrix}
     \ ,
\end{equation}
where $\theta$ is the angle between $\hat{\mathbf{u}}$ and $\hat{\mathbf{v}}$ \citep{2004SoPh..219....3D, riley2006mathematical}. 

Following this re-projection, the angle of the magnetic field (and hence the fibril) with respect to the line-of-sight of the observer was measured along a magnetic field line above each of the cross cut locations, which are shown in Figure~\ref{fig:fibril}. This line was chosen by seeding magnetic field lines at each height above and below all cross cut locations and selecting a field line that passes through all cross cut locations and most closely matches the average fibril axis. This allows for the re-projection of all LOS and POS measurements into the Fibril Axis Frame (FAF). As the name suggests, the FAF-$z$-axis is oriented along the axis of the fibril, away from the sunspot (blue arrow in Figure~{\ref{fig:fib_cart}}). The FAF-$x$-axis is perpendicular to the FAF-$z$-axis and parallel to the solar surface (red arrow in Figure~{\ref{fig:fib_cart}}). The FAF-$y$-axis is mutually perpendicular to both the FAF-$x/z$ axes (green arrow in Figure~{\ref{fig:fib_cart}}). The complete setup is visualised in Figure~\ref{fig:fib_cart}.

\section{Results}
\label{sec:res}

\begin{figure}[t]
\centering
\includegraphics[trim=0mm 20mm 0mm 25mm, clip, width=\columnwidth, angle=0]{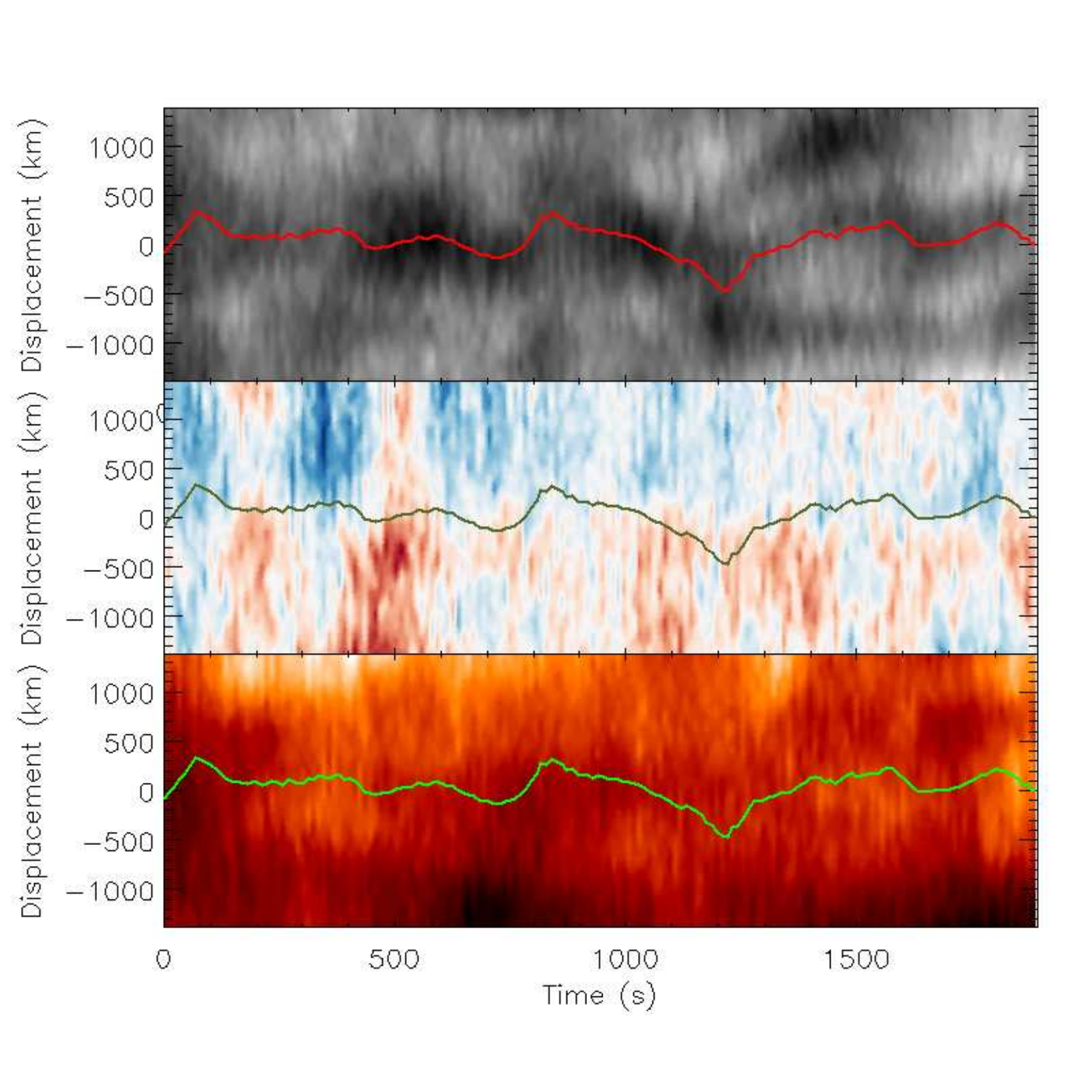}
 \caption{Time distance diagrams illustrating the data analysis of a fibril oscillation, with a displacement of $0$~km corresponding to the average fibril axis. The top panel shows line core intensity, with the overplotted red line representing the calculated central position of the fibril as it undergoes FAF-$x$ transverse oscillations. This same line is overplotted on the middle and bottom panels. The middle panel shows FAF-$y$ velocity, and the bottom shows the H-alpha line width. 
\label{fig:td}}
\end{figure}

\begin{figure}[t]
\centering
\includegraphics[trim=0mm 0mm 0mm 0mm, clip, width=\columnwidth, angle=0]{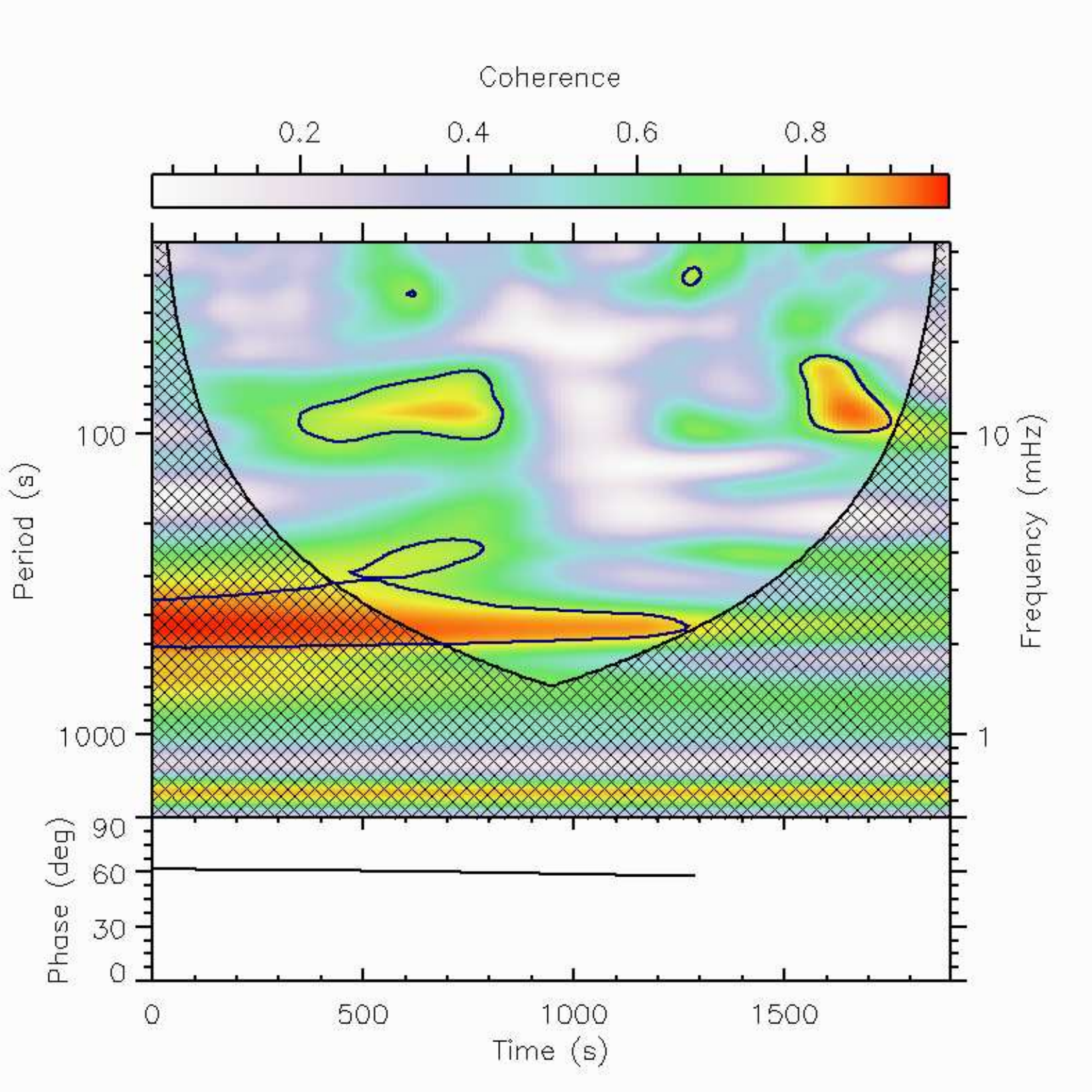}
 \caption{A cross-wavelet coherence spectrum between the FAF-$x$ displacement and the FAF-$y$ velocity measured at the cut location denoted by the green cross in Figure~\ref{fig:fibril}. The cone of influence is shown by the hashed region, the contour is at a $0.8$ coherence level. The lower panel shows the phase angle measured at the dominant frequency ($\sim430$~s) as a function of time.
\label{fig:cross_vlos}}
\end{figure}

In the following subsections, the data analysis techniques detailed in Section~\ref{sec:ana} are demonstrated through thorough analysis of a single fibril, shown by the white line in Figure~\ref{fig:fibril}. We detail the three-dimensional kink-like motion of the fibril in the Fibril Axis Frame (Section~\ref{sub:kink}), examine the trends of the H$\alpha$ line width both as a function of distance and of time (Section~\ref{sub:width_wobble}), and use these observed oscillatory properties to estimate the energy flux carried by the fibril's oscillations (Section~\ref{sub:flux}). 

\subsection{Kink Oscillations}
\label{sub:kink}

Transverse oscillations were found to be near ubiquitous within the super-penumbral fibrils identified. An example fibril is shown in Figure~\ref{fig:fibril} with the locations of the cross cuts taken denoted by coloured crosses and the fibril axis denoted by the white line. The magnetic field angles with respect to the solar surface at the heights and distances from the centre of the sunspot considered within this study were found to be within the range $10-40\degree$, in agreement with \citet{2013ApJ...779..168J, 2021RSPTA.37900183M}. The TD diagrams shown in Figure~\ref{fig:td} were taken at the green cross shown in Figure~\ref{fig:fibril}. The magnetic field angle of this fibril at the green cross was found to be $22\degree$ with respect to the solar surface and $-4\degree$ with respect to plane-of-sky. As a result all line-of-sight velocity measurements were adjusted by an appropriate factor, recasting $v_{\text{LOS}}$ to a new $v_{y}$ in the FAF, as detailed in Section~\ref{sub:reproj}.

\begin{figure*}[t]
\centering
\includegraphics[trim=0mm 0mm 0mm 0mm, clip, width=\textwidth, angle=0]{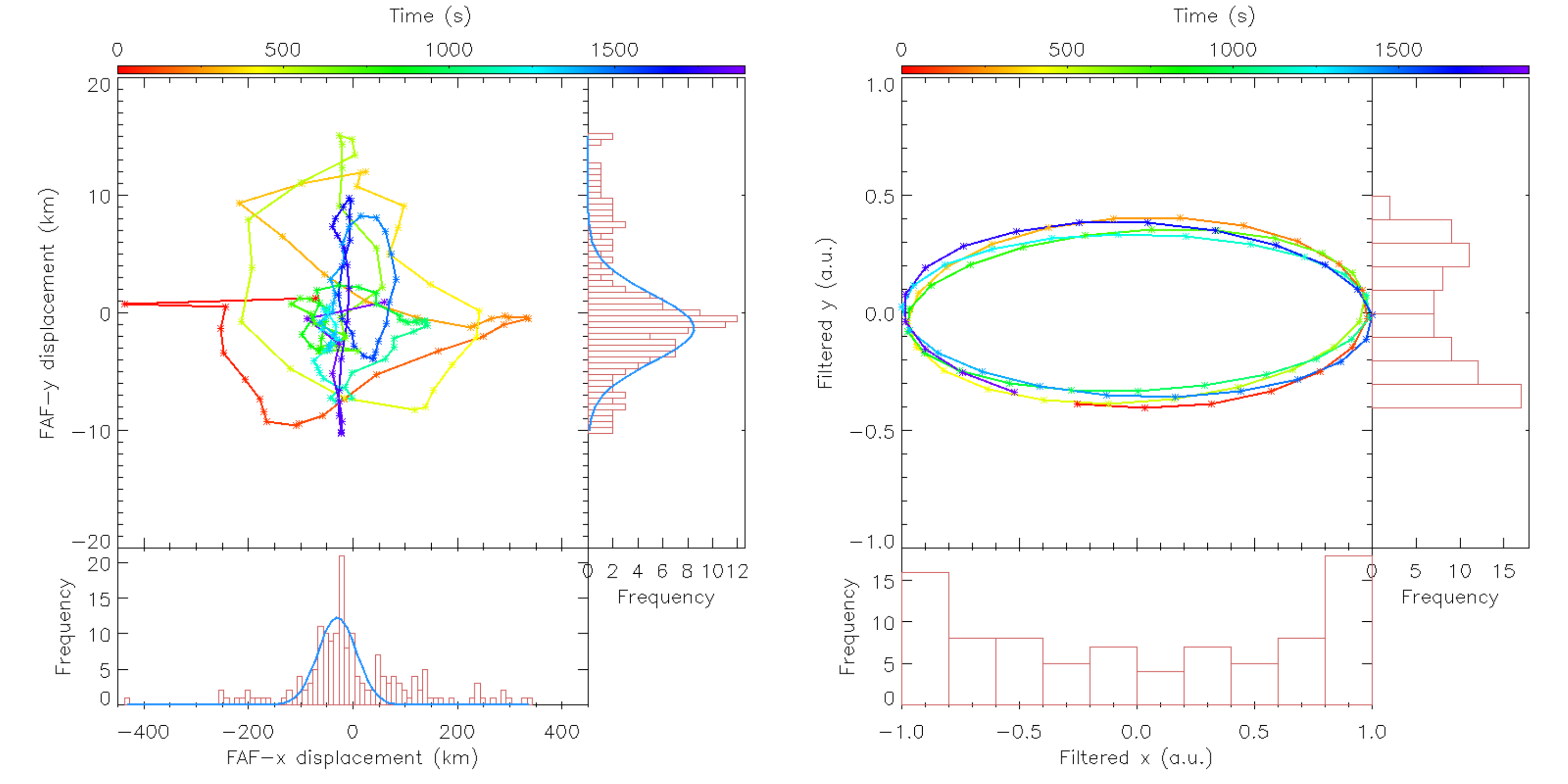}
 \caption{Hodograms representing the path that the fibril oscillation takes in the FAF-$x$ and $y$ directions from and end-on view i.e. the path it would trace out on the grey $x,y$ plane marked in Figure~\ref{fig:fib_cart}. The left panel shows the raw data and the right panel shows the Fourier filtered data for the $430$~s oscillations under consideration. Time is represented by the colours indicated by the colour bar at the top. The histograms on the bottom and right of each panel show how long the fibril spends at each of the FAF-$x$ and $y$ positions respectively. These measurements were taken at the location of the orange cross-cut in Figure~\ref{fig:fibril}. It should be noted that the amplitude of the oscillations in the FAF-$y$ direction shown in the left panel are over an order of magnitude smaller than those in the FAF-$x$ direction.
\label{fig:hodogram}}
\end{figure*}

Wavelet analysis \citep{1998BAMS...79...61T} of the transverse oscillations at each of the cross cut locations show the same dominant period of $\sim430$~s. This is consistent with the median and modal periods, of $570$~s and $330$~s respectively, found by \citet{2021RSPTA.37900183M} in super-penumbral fibrils. This is also consistent with the work of \citet{2013ApJ...779..168J} who found that longer period ($>180$~s) waves were dominant in the sunspot penumbra and beyond, due to the effect of the magnetic field inclination on the acoustic cutoff frequency \citep{1977A&A....55..239B}. Oscillations with a similar periodicity were also identified in the $v_{y}$ oscillations. This then allowed for the use of cross-wavelet analysis to investigate the coherence of these oscillations as well as the phase lag between them. A cross-wavelet coherence spectrum is shown in Figure~\ref{fig:cross_vlos} comparing the transverse displacement ($s_{x}$) and $v_{y}$ taken at the cross cut denoted by the green cross in Figure~\ref{fig:fibril}. The $430$~s oscillations are shown to be coherent and there is a phase lag of $60\degree$ between $v_{y}$ and $s_{x}$, which is consistent at each cross cut along the fibril. This phase lag also remains constant in time, as can be seen in the lower panel of Figure~\ref{fig:cross_vlos}. Physically this phase lag implies an elliptically polarised kink wave, with its semi-major and semi-minor axes neither parallel or perpendicular to our FAF-$x,y$ axes \citep{Lissajous1857}. Through further analysis, the velocity amplitude of the transverse displacement oscillations was found to be $v_{\text{amp},x}=1.6$~km{\,}s$^{-1}$, whilst the amplitude of the $v_{y}$ oscillations was found to be $v_{\text{amp},y}=0.3$~km{\,}s$^{-1}$. This larger amplitude in the transverse direction is likely due to the suppression of FAF-$y$ oscillations by the density stratification of the solar atmosphere. If the FAF-$x$ and FAF-$y$ oscillations were treated as two separate kink oscillations, then there would be approximately $30$ times more energy flux carried by the FAF-$x$ oscillation due to the energy flux being proportional to the square of the velocity amplitude \citep{VanDoorsselaere2014}. 

In order to better illustrate the polarisation of the transverse oscillation, a hodogram (or phase portrait) is presented in Figure~\ref{fig:hodogram}. This was constructed through integration of $v_{y}$ in order to calculate the FAF-$y$ displacement, $s_{y}$. A hodogram was then produced following a similar method to that used by \citet{2023NatCo..14.5298Z}. The histogram distributions are shown at the bottom and on the right of each hodogram. The left panel shows the unfiltered data and the histograms shown are fitted with Gaussian distributions. The right panel has the data Fourier filtered for the $430$~s oscillation under investigation. The $x$-axis range in the left panel is $[-450,450]$~km, whilst the $y$-range is $[-20,20]$~km. The units shown in the right panel are arbitrary. The bimodal nature of the histograms helps to confirm the elliptical nature of the oscillation, rather than a oblique-linear polarisation \citep{2023NatCo..14.5298Z}.

\begin{figure}[t]
\centering
\includegraphics[trim=0mm 0mm 0mm 0mm, clip, width=\columnwidth, angle=0]{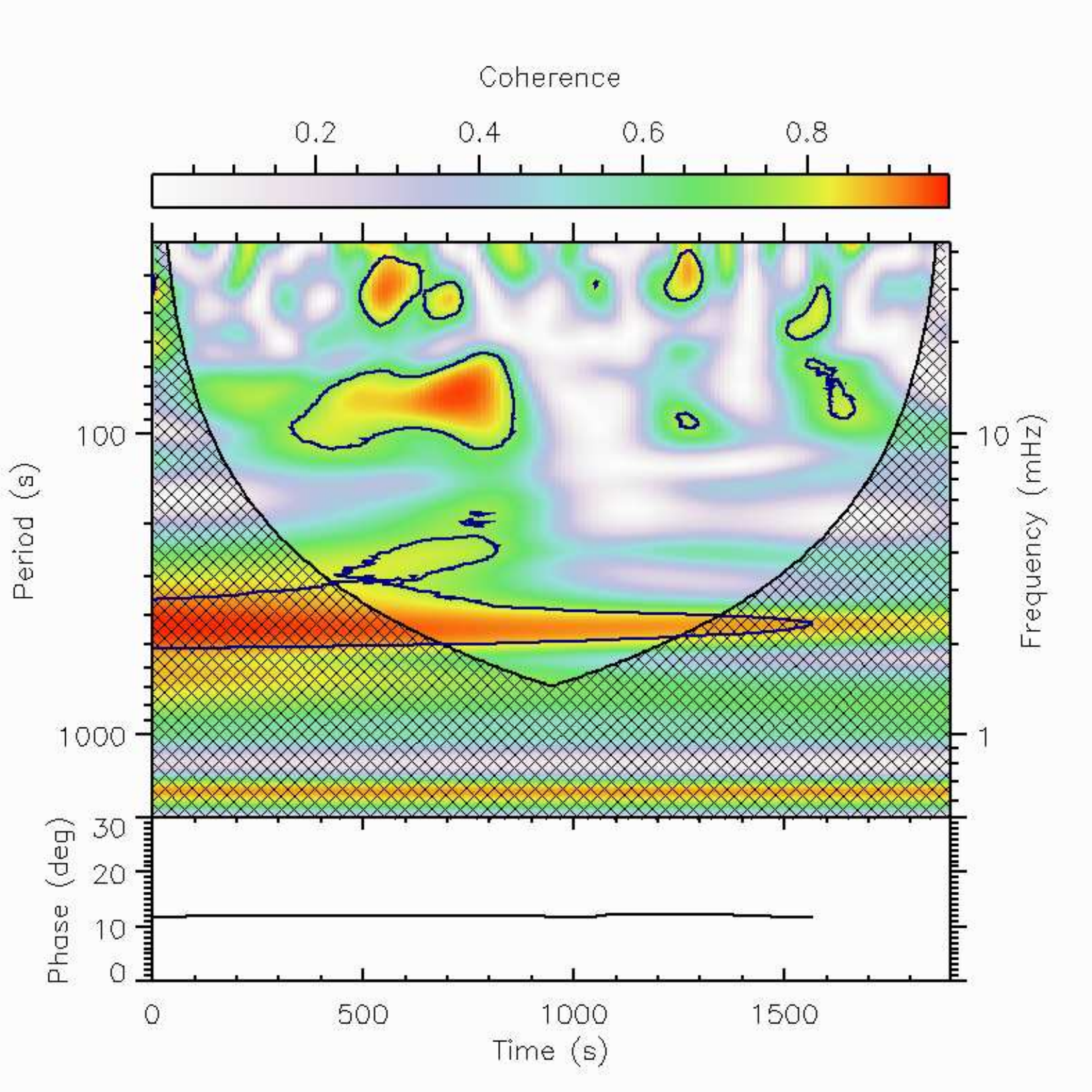}
 \caption{The same as Figure~\ref{fig:cross_vlos} but comparing the FAF-$x$ oscillations measured at the locations of the orange and green crosses in Figure~\ref{fig:fibril}. The lower panel shows the phase angle measured at the dominant frequency ($\sim430$~s) as a function of time.
\label{fig:cross_ph}}
\end{figure}

The phase speed, $v_{\text{ph},z}$, of the oscillations along the FAF-$z$ direction was calculated using phase lag analysis \citep[e.g.,][]{1997ApJ...474L..43V}. Similar cross wavelet procedures were employed when comparing the $s_{x}$ oscillations at different cut locations along the fibril length. The cross-wavelet coherence spectrum between $s_{x}$ measured at the orange and green cross cuts is shown in Figure~\ref{fig:cross_ph}. The phase lag, $\phi$, was found to be $12\degree$ and was constant with time, as can be seen in the lower panel of Figure~\ref{fig:cross_ph}. This, combined with distance along the FAF-$z$ axis between the cut locations being $d_{z} =1200$~km and knowledge of the period, $\mathcal{T}$, allows for the calculation \citep[e.g.,][]{1977SoPh...52..283M,2015ApJ...806..132G} of the phase speed as,
\begin{equation}
\label{eqn:phase_speed}
    v_{\text{ph},z} = \frac{360 d_{z}}{\mathcal{T}\phi} \ .
\end{equation}
It was found that the oscillation was travelling along the fibril away from the sunspot centre and $v_{\text{ph},z}$ was calculated for each pair of adjacent cross cuts along the fibril using the same cross wavelet analysis shown for the green and orange cross cuts in Figure~\ref{fig:cross_ph}. The phase speed of the transverse oscillations of the fibril were found to be constant along the length of the fibril at $69\pm4$~km{\,}s$^{-1}$. This lack of evidence for a change in propagation speed as a function of distance is in keeping with the findings of both \citet{2017A&A...607A..46M} and \citet{2021RSPTA.37900183M}. This value for the phase speed is in good agreement with previous values for kink waves found in quiet Sun and internetwork fibrillar structures found by \citet{2012NatCo...3.1315M} and \citet{2013ApJ...779...82K}. However, it is towards the higher end of the distribution of propagation speeds in super-penumbral fibrils calculated by \citet{2021RSPTA.37900183M}, although this is likely due to their longer time cadence of $30$~s. This phase velocity is much larger than that found in running penumbral waves of $5-20$~km{\,}s$^{-1}$ \citep{1997ApJ...478..814B, 2013ApJ...779..168J}.

\subsection{Line Width Trend and Oscillation}
\label{sub:width_wobble}

The line width is seen to increase as we move away from the sunspot in all directions, as in the right panel of Figure~\ref{fig:FOV}. In order to account for this when considering the fibril alone, the average line width at each distance away from the sunspot was calculated over an arc of $\pi/2$ radians centred on the fibril axis. This average line width decreases at a rate of $0.009$~\AA/Mm with distance from the sunspot and shows no significant variation with time. This average value was then subtracted from the line width calculated at each cross cut resulting in new quantity henceforth referred to as global trend subtracted (GTS) line width. This subtraction of the background trend was performed in order to isolate local effects within the fibril from general trends in line width seen in the right panel of Figure~\ref{fig:FOV} and ensures that the data plotted represents line width variations intrinsically linked to the fibril itself. The GTS line width increases at a rate of $0.017$~\AA/Mm with distance along the fibril.

\begin{figure}[t]
\centering
\includegraphics[trim=0mm 0mm 0mm 0mm, clip, width=\columnwidth, angle=0]{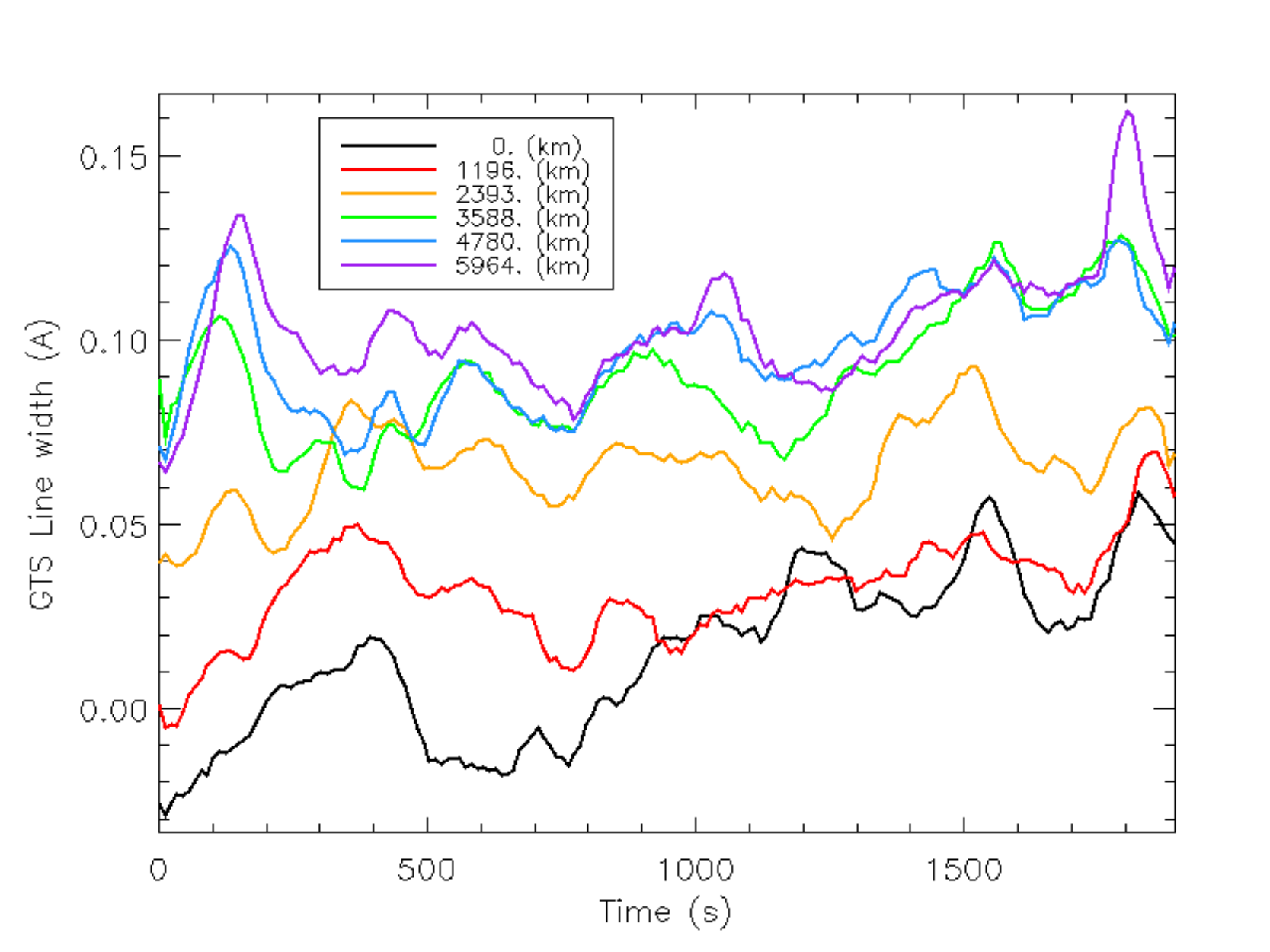}
 \caption{The global trend subtracted (GTS) H$\alpha$ line width calculated at different locations along the fibril. The colours correspond to those shown in Figure~\ref{fig:fibril}. 
\label{fig:temp}}
\end{figure}

This new GTS line width at each of the cross cut locations along the fibril is shown in Figure~\ref{fig:temp}, with the colour of each line corresponding to the locations denoted by the same colour in Figure~\ref{fig:fibril}. The GTS line width can be seen to be increasing as a function of distance along the fibril. There is also a general trend of increasing GTS line width with time. The oscillations that can be seen in the GTS line width are believed to be due to an independent sausage-type oscillation coexisting within the fibril \citep{1982SoPh...75....3S, 1983SoPh...88..179E, 2012ApJ...744L...5J, 2012NatCo...3.1315M, 2018ApJ...853...61S}. 

\begin{figure}[t]
\centering
\includegraphics[trim=0mm 0mm 0mm 0mm, clip, width=\columnwidth, angle=0]{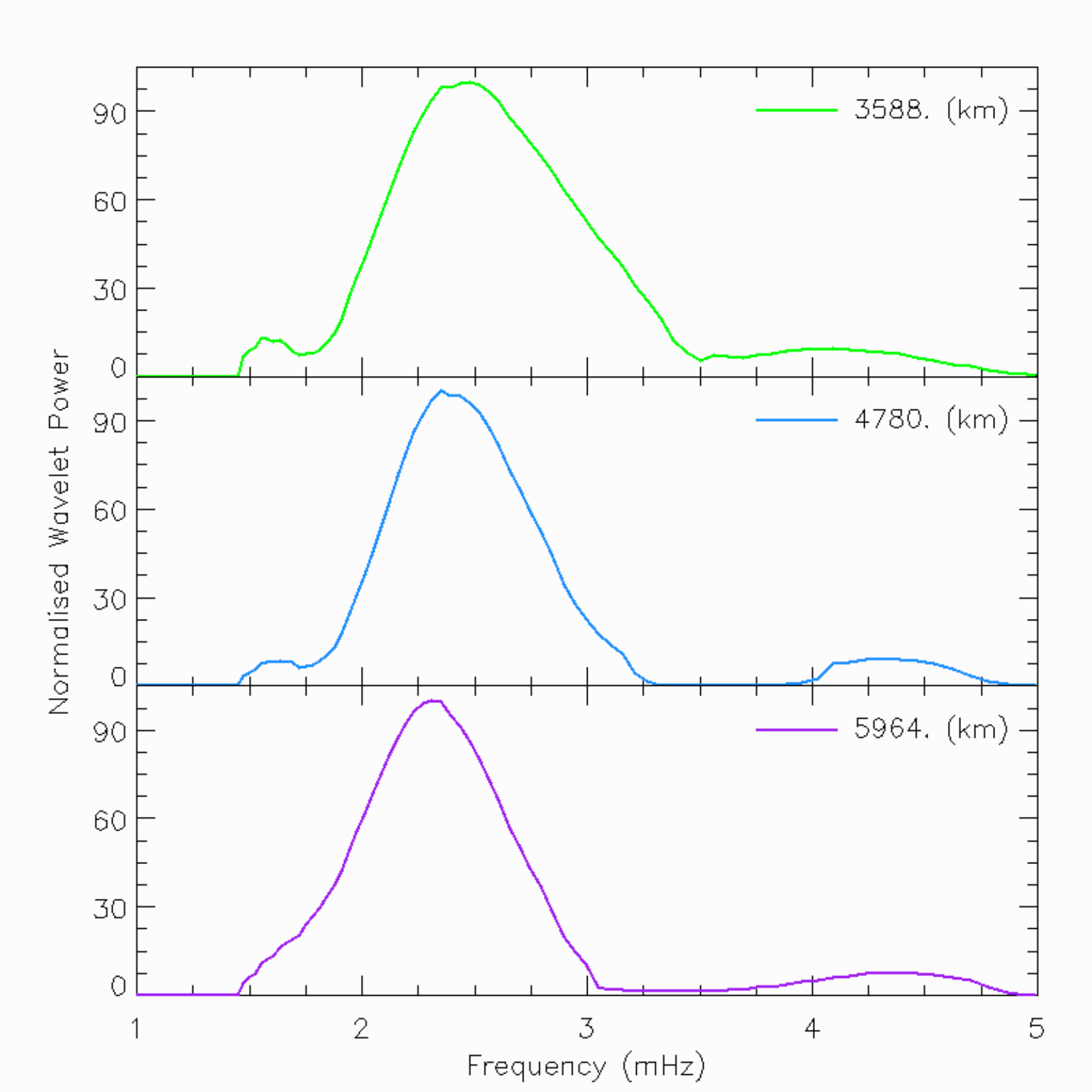}
 \caption{Wavelet power spectra calculated from the time series of H$\alpha$ GTS line width shown in Figure~\ref{fig:temp}. The colours correspond to the locations of the cross cuts used for measurement shown in Figure~\ref{fig:fibril}.  Each power spectrum shows a significant power at approximately $410$~s.
\label{fig:wavelet}}
\end{figure}

This periodicity was analysed by computing power spectra of the GTS line width taken at the green, blue, and purple cross cut locations in Figure~\ref{fig:fibril} using a time-integral of wavelet power excluding regions influenced by cone-of-influence and only for those above the confidence level, resulting in a power-weighted frequency distribution \citep{1998BAMS...79...61T}. These power spectra are shown in Figure~\ref{fig:wavelet}. A clear peak can be seen in the periodicity of each of these GTS line width time series, located at a frequency consistent with a period of $410$~s. This periodicity is extremely similar to that measured for the kink oscillation, $430$~s. Further, phase-lag analysis \citep[e.g.,][]{1997ApJ...474L..43V} was undertaken and a phase lag of $8\degree$ was found between both sets of adjacent cross cuts (green and blue; blue and purple). This allows the calculation of the phase velocity of this oscillation using Equation~\ref{eqn:phase_speed} as $130$~km~s$^{-1}$, propagating away from the sunspot. This is notably larger than the previously calculated phase speed of the kink oscillation, at $69\pm4$~km~s$^{-1}$. This is not unexpected, both the kink and sausage waves have phase speeds well above the typical chromospheric sound speed, identifying them as fast waves \citep{2012NatCo...3.1315M}. In this regime, fast sausage waves have greater phase speeds than fast kink waves \citep{1983SoPh...88..179E}, indicating the GTS line width oscillations may be the signature of axisymmetric sausage modes.

If this was to be interpreted as a sausage mode detected concurrently with a kink mode within a fibril, then the similar periods of $410$~s and $430$~s respectively would be consistent with the findings of \citet{2012NatCo...3.1315M}, who found that their detected kink and sausage oscillations had periods of $232\pm8$~s and $197\pm8$~s respectively. In order to conclusively associate the GTS line width oscillations with a sausage oscillation, it was desired to observe a contemporaneous and co-periodic oscillation of the width of the magnetic cylinder, in this case our fibril \citep{1983SoPh...88..179E}. However, no such oscillations were detected in our dataset when following the methodology of \citet{2012NatCo...3.1315M}. This is likely due to any changes in fibril width being below the detection limit governed by the spatial resolution. Due to the uncertainty in the amplitude, we are unable to conclusively calculate the energy flux of the embedded sausage wave. Nevertheless, we present evidence of sausage-mode oscillations existing within the fibrillar structure, hence providing an interesting avenue of further investigation for future studies.

\vspace{5mm}

\subsection{Kink Energy Flux Estimation}
\label{sub:flux}

By utilising a model for the density of the fibril, following the variation with height above the solar surface according to \citet{1986ApJ...306..284M}, and by treating the displacement oscillations as a kink mode with velocity amplitude equal to the quadrature sum of the FAF-$x$ and $y$ velocity amplitudes, the energy flux, $F$, of the oscillation can be calculated as,
\begin{equation}
    F \approx f\frac{1}{2}\rho_{i}v_{\text{amp}}^{2}v_{\text{ph},z} \ ,    \label{eqn:flux}
\end{equation}
where $\rho_{i}$ is the density inside the fibril, and $f$ is the filling factor that was taken as $5\%$ to remain consistent with previous literature values \citep[e.g.,][]{VanDoorsselaere2014, 2022ApJ...930..129B, 2022A&A...659A..73M}.

\begin{figure}[!t]
\centering
\includegraphics[trim=0mm 0mm 0mm 0mm, clip, width=\columnwidth, angle=0]{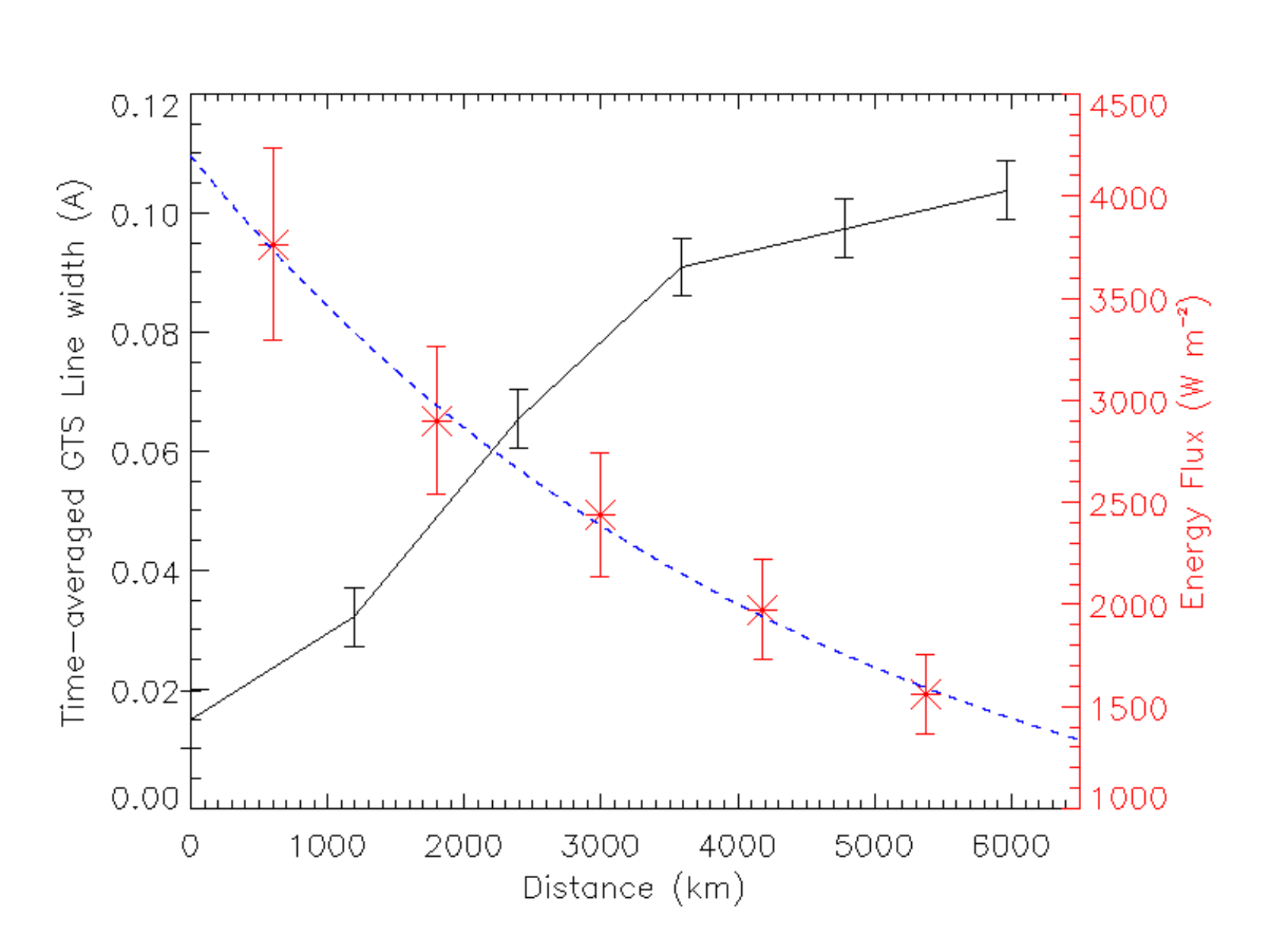}
 \caption{A comparison of time-averaged GTS line width at each distance, shown in black; with the energy flux carried by the kink wave calculated using equation~\ref{eqn:flux}, shown in red. An exponential decay fit to the energy flux is shown by the blue dashed line.
\label{fig:flux}}
\end{figure}

The time-averaged GTS line width and the energy flux are presented as a function of distance in Figure~\ref{fig:flux}. The line width increases at an average rate of $0.017$~\AA/Mm. As noted in Section~\ref{sub:line}, care must be taken when using the relationship presented in Equation~\ref{eqn:tempwidth} as different gradients have been found for different H$\alpha$ datasets \citep{2023FrASS...978405T}. However, assuming that Equation~\ref{eqn:tempwidth} holds for the purposes of temperature comparison, we arrive at a temperature increase rate of approximately $250$~K/Mm. Although, as mentioned previously, non-thermal effects can contribute to line width, relevant simulations have found that temperature increase is the dominant source of line broadening when compared to small scale turbulence \citep[e.g.][]{2017ApJ...836..219A}.

The energy flux is seen to decrease at an average rate of $460$~W{\,}m$^{-2}$/Mm. This increase in GTS line width, coupled with a decrease in the energy flux carried by the kink waves, appears to be a signature of MHD wave damping in the form of plasma thermalisation.

As is shown by the blue dashed line in Figure~\ref{fig:flux}, the energy flux decay can be approximated by an exponential decay as a function of distance, with a scale height of $4.6$~Mm. A linear damping mechanism would correspond to such an exponential decay, whereas a non-linear damping mechanism would result in a different decay profile \citep{2023arXiv230802217H}. However, it is difficult to infer the true nature of the relationship with only five data points. Due to energy flux being proportional to the square of velocity amplitude in Equation~\ref{eqn:flux}, this energy scale height of $4.6$~Mm implies a damping length of $9.2$~Mm.

\section{Theoretical Mechanisms for Wave Damping}
\label{sec:modelling}

Circularly polarised propagating kink waves have been observed before in chromospheric magnetic elements \citep{2017ApJ...840...19S}. These circularly/elliptically polarised kink modes should not be unexpected as, even if all kink modes are initially linearly polarised, circular polarisation will evolve in magnetic structures with twisted magnetic fields or rotational flows \citep{2015A&A...580A..57R}. Further, expansion of the magnetic feature hosting linearly polarised kink modes can result in elliptical polarisation arising \citep[e.g.,][]{2015A&A...581A...8R}. Another excitation mechanism for elliptically polarised kink waves is the combination of two or more kink pulses with different orientations and amplitudes are superimposed within the same magnetic flux tube \citep{2008ApJ...683L..91Z,2017ApJ...840...19S}. \citet{2022A&A...659A..73M} examine the dynamics of circularly polarised standing kink waves and their damping. They find that through the damping of these circularly polarised waves, elliptical polarisation again develops.

There are many proposed mechanisms for the damping of kink (or Alfv\'{e}nic) waves \citep[see the recent review by][and references therein]{2023RvMPP...7...17M}. By treating fibrils as magnetic cylinders embedded within the solar chromosphere, this plasma inhomogeneity results in more effective damping than Alfv\'{e}n waves would otherwise experience in a uniform medium. It should be noted that this damping does not necessarily imply the deposition of wave energy into the plasma as heating. Here we explore the different theoretical mechanisms for wave damping and attempt to interpret how these would affect the kink oscillation of a fibril both independently and through their interplay.
\vspace{1em}

\subsection{Resonant Absorption}
\label{sub:resabs}

Within our magnetic cylinder approximation of an oscillating fibril, resonant absorption relies on the existence of a continuous variation of the Alfv\'{e}n speed perpendicular to the magnetic field (as opposed to a discrete step function at the boundary of the cylinder). This will be assumed to be achieved by a variation of density, $\rho$, for further calculations in this subsection \citep{2011SSRv..158..289G}. Resonant absorption as a wave damping mechanism relies on the energy of a global MHD wave (in this case a kink wave) being transferred into small-scale resonant Alfv\'{e}n waves that then lead to dissipation. These small-scale waves may also result in line broadening \citep{2005pps..book.....G}.

The oscillations analysed in Section~\ref{sec:res} lie firmly within the long wavelength regime $(R/\lambda \ll 1)$, where $R$ is the radius of the fibril and $\lambda$ is the wavelength of the kink wave \citep{2010A&A...524A..23T}. This allows for the use of the Terradas–Goossens–Verth (TGV) relation, given by

\begin{equation}
\label{eqn:TGV}
    L_{D} = v_{\text{ph},z} \frac{2}{\pi} \frac{1}{m} \frac{R}{l} \frac{\rho_{i}+\rho_{e}}{\rho_{i}-\rho_{e}} \frac{1}{\nu} \ ,
\end{equation}
where $R$ is the radius of the fibril, $l$ is the width of the inhomogeneous surface layer, $\rho_{i,e}$ are the internal and external densities respectively, and $\nu$ is the frequency of the oscillation \citep{2010ApJ...718L.102V, 2019ApJ...876..106T}. 

As the exact density contrast, $\zeta=\rho_{i}/\rho_{e}$, is unknown, a range of values are considered. The factor $(\rho_{i}+\rho_{e})/(\rho_{i}-\rho_{e})$ varies between $2$ and $1$ for a density contrast range of $\zeta=[3,\infty)$. This allows for the consideration of all sensible density contrast values. The width of the boundary layer is also unknown, but a range of values $l/R=[1/3,1/2]$ is considered \citep[e.g.,][]{2016A&A...595A..81M, 2021ApJ...923..225M}.

Using this relationship, assuming a pure kink mode $(m=1)$, a range of damping lengths between $38-113$~Mm due to resonant absorption is calculated. The most effective damping (and thus the shortest damping length) corresponds to the case $\zeta \rightarrow \infty$ and $l/R=1/2$. Through the recasting of this damping length as a damping time, valid for cases of weak damping \citep[e.g.,][]{2021ApJ...923..225M}, the damping time due to resonant absorption can be calculated as $\tau_{R} = 550-1650$~s.

The calculated damping length due to resonant absorption far exceeds the observed damping length, even in the most extreme case, calculated from the relationship shown in Figure~\ref{fig:flux} as $9.2$~Mm. It should be noted however, that Equation~\ref{eqn:TGV} is formulated under the assumption of weak damping \citep{2004ApJ...606.1223V, 2013ApJ...777..158S}. This indicates that resonant absorption of pure kink modes is not responsible for the damping reported in this study. However, as can be seen from Equation~\ref{eqn:TGV}, higher order fluting modes $(m>1)$ damp due to resonant absorption over shorter length scales than pure kink modes. Pure fluting modes in a magnetic cylinder are surface-like and in smooth boundary cases (such as our $l/R =[1/3,1/2]$ assumption), these modes are likely not able to be supported \citep{2017ApJ...850..114S}. However, numerical simulations along with analytic theory suggest that kink motions can couple with fluting modes non-linearly, resulting in a fraction of the kink mode energy being converted to fluting modes and the production of `mixed-modes', so this route of wave damping via mode mode conversion cannot be discounted \citep{2010PhPl...17h2108R, 2017SoPh..292..111R, 2018ApJ...853...35T}.

The damping due to ion-neutral collisions (Cowling's diffusion) has not been considered here, as it has been found through theoretical studies to be negligible when compared to the damping due to resonant absorption \citep{2011ApJ...726..102S, 2012A&A...537A..84S}.

\subsection{Uniturbulence}
\label{sub:uni}

In a uniform medium, Alfv\'{e}n wave turbulence is generally understood to arise from the non-linear interaction of counter-propagating Alfv\'{e}n waves \citep{1964SvA.....7..566I, 1965PhFl....8.1385K}. This is typically understood through use of the Els\"{a}sser formalism. In an incompressible, homogeneous plasma, pure Alfv\'{e}n waves are described by either one of the Els\"{a}sser variables, with $z^{-}$ corresponding to propagation parallel to the magnetic field, and $z^{+}$ to anti-parallel propagation. A necessary condition for the formation of turbulence is that both $z^{\pm}$ are non-zero, hence the need for counter-propagating waves.

However, the solar atmosphere is not well described by an incompressible, homogeneous plasma and so further refinements are required. Perpendicular inhomogeneities with respect to the magnetic field can also give rise to Alfv\'{e}nic wave turbulence, such as in the case of a magnetic cylinder interpretation of a fibril. This is due to the inhomogeneity allowing for both $z^{\pm}$ to be non- zero for an kink wave propagating in a single direction. Unidirectionally propagating kink waves can be damped by so called `uniturbulence' \citep{2017NatSR...714820M, 2019ApJ...882...50M}, which is a non-linear damping mechanism with an energy cascade damping time given by \citet{2020ApJ...899..100V, 2021ApJ...920..162V} as,
\begin{equation}
\label{eqn:unidamp}
    \tau_{U} = 2\sqrt{5\pi} \frac{2R}{v_{\text{amp}}} \frac{\rho_{i}+\rho_{e}}{\rho_{i}-\rho_{e}} \ .
\end{equation}
The non-linearity of this damping mechanism is made clear by the influence of the velocity amplitude, $v_{\text{amp}}$, in Equation~\ref{eqn:unidamp}. This interpretation of $\tau_{U}$ as a damping time relies on the assumption of an exponential decay profile due to this non-linear damping mechanism, something which cannot hold, but is an approximation made for comparison with linear damping mechanisms \citep{2021ApJ...920..162V}.

By utilising the same range of density contrasts, $\zeta=[3,\infty)$, as in Section~\ref{sub:resabs}'s damping length calculation, and our largest calculated velocity amplitude, $v_{\text{amp}}=1.9$~km~s$^{-1}$, we are able to calculate an damping time range of $\tau_{U}=1400-2800$~s. This is noticeably larger than the $\tau_{R}$ of $550-1650$~s previously calculated. This is in good agreement with the study by \citet{2021ApJ...923..225M} which found that $\tau_{R}/\tau_{U}$ is typically less than 1 for propagating kink modes. It should also be noted that the strength of damping due to uniturbulence decreases for smaller velocity amplitudes, decreasing in its effectiveness by almost a factor of $2$ for the smallest calculated valocity amplitude of $v_{\text{amp}}=1.2$~km~s$^{-1}$.

\subsection{Combination of Damping Mechanisms}
\label{sub:comb}

If the contribution of two damping mechanisms is small and independent, it is possible to combine their effects following the methodology of \citet{2021ApJ...910...58V} and \citet{2021ApJ...923..225M}. This assumption of independence may not be entirely justified due to the effects of resonant absorption on the development on turbulence, but will prove useful as a first order approximation \citep{2019FrP.....7...85A}. It also assumes that the damping effects of both mechanisms result in exponential damping profiles, something which does not hold for non-linear damping, but due to the relatively weak effect of the non-linear damping, this should not invalidate its results \citep{2021ApJ...923..225M}. This approach combines the damping times due to resonant absorption (see Section~\ref{sub:resabs}) and the dominant non-linear damping mechanism of uniturbulence (see Section~\ref{sub:uni}) using the following\footnote{It should be noted that previous works \citep[e.g.][]{2021ApJ...910...58V, 2021ApJ...923..225M} refer to this relationship as a ``harmonic average'', however, this relationship is distinct from the more common definition of a harmonic mean by a factor of $2$.} to calculate the overall damping time, $\tau_{T}$, as follows

\begin{equation}
    \label{eqn:t_tot}
    \frac{1}{\tau_{T}} = \frac{1}{\tau_{R}} + \frac{1}{\tau_{U}} \ .
\end{equation}

Following this approach, we arrive at $\tau_{T} = 400-1050$~s. Recasting this as a damping length gives a value of $27-71$~Mm. Although this is shorter than the damping length derived from resonant absorption and uniturbulence individually, even in the most extreme case it is a factor of $3$ longer than the damping length of $9.2$~Mm derived from the relationship shown in Figure~\ref{fig:flux}. This is likely due to the more complex interplay between different damping mechanisms; other damping mechanisms not considered; or different values of density contrast ($\zeta$) and boundary layer thickness ($l$) being the case for the fibril considered.

\section{Conclusions} 
\label{sec:conc}

Here we provide an in-depth case study of a transverse fibril oscillation and provide numerous techniques to help analyse its motion. There are a large number of previous studies of transverse waves within fibrillar structures. Through magnetic field modelling via a NLFFF extrapolation and consideration of the effect of the solar $\mu$ angle, the transverse oscillation characteristics have been re-projected into the Fibril Axis Frame (FAF). This has allowed for the alleviation of projection effects that would be suffered when performing a standard helioprojective study. It is anticipated that this technique can further inform future works on the topic, helping to move away from a simple line-of-sight and plane-of-sky analysis. The in-depth discussion of this methodology should allow for its use in larger statistical studies of the oscillations of fibrillar structures, without the major concern of projection effects.  

This has allowed an elliptically polarised kink wave of a super-penumbral fibril to be resolved and its propagation away from the sunspot centre to be characterised, with a phase lag of $60\degree$ between the FAF-$x$ and $y$ motions and a period of $430$~s. This represents the first observation of an elliptically polarised kink wave within a chromospheric fibril using properly translated motions. These observed and calculated oscillation parameters have facilitated the calculation of the energy flux carried by this kink wave as it propagates in the FAF-$z$ direction at a phase velocity of $v_{\text{ph},z}=69\pm4$~km~s$^{-1}$ and its associated damping length of $L_{D}=9.2$~Mm. Further, through the use of a novel `super-sech' function for the fitting of the H$\alpha$ line profile, an increase in the GTS line width has been tracked showing an increase with distance along the fibril in the FAF-$z$ direction. This is an indicator of wave damping occurring alongside plasma thermalisation, although further work is necessary to ascertain whether the two processes are linked through energy dissipation.

Different mechanisms have been explored for the damping of the energy flux carried by the wave, with resonant absorption and uniturbulence considered as the dominant linear and non-linear mechanisms respectively. However, through a first-order approximation of their combined effect, we arrive at a shortest theoretical damping length of $27$~Mm. This is $3$ times longer than the observed damping length, even for the most extreme combination of parameters considered, demonstrating that further analysis of potential damping mechanisms and their interplay is warranted. 

As mentioned previously, it is anticipated that the techniques and methodology here can be used to inform future studies of the ubiquitous transverse waves found within fibrillar structures. The H$\alpha$ profile fitting and the FAF re-projection code will be made publicly available with the upcoming release of WaLSAtools \citep{2023LRSP...20....1J}.

\section*{Acknowledgements}
\begin{acknowledgments}
WB, DBJ and TD acknowledge support from the Leverhulme Trust via the Research Project Grant RPG-2019-371.
DBJ and SDTG wish to thank the UK Science and Technology Facilities Council (STFC) for the consolidated grants ST/T00021X/1 and ST/X000923/1.
DBJ and SDTG also acknowledge funding from the UK Space Agency via the National Space Technology Programme (grant SSc-009).
AH is supported by STFC Research Grant No. ST/V000659/1.
TVD received financial support from the Flemish Government under the long-term structural Methusalem funding program, project SOUL: Stellar evolution in full glory, grant METH/24/012 at KU Leuven, the DynaSun project (number 101131534 of HORIZON-MSCA-2022-SE-01), and also a Senior Research Project (G088021N) of the FWO Vlaanderen.
TW acknowledges DLR grant 50OC2301.
The Dunn Solar Telescope at Sacramento Peak/NM was operated by the National Solar Observatory (NSO). NSO is operated by the Association of Universities for Research in Astronomy (AURA), Inc., under cooperative agreement with the National Science Foundation (NSF). 
Finally, we wish to acknowledge scientific discussions with the Waves in the Lower Solar Atmosphere (WaLSA; \href{https://www.WaLSA.team}{https://www.WaLSA.team}) team, which has been supported by the Research Council of Norway (project no. 262622), The Royal Society \citep[award no. Hooke18b/SCTM;][]{2021RSPTA.37900169J}, and the International Space Science Institute (ISSI Team~502).  
\end{acknowledgments}

%

\vspace{5mm}
\facilities{DST \citep[IBIS;][]{2006SoPh..236..415C}, HMI \citep[HMI;][]{2012SoPh..275..327S}}





\bibliography{apjbib}{}
\bibliographystyle{aasjournal}

\end{document}